\documentstyle[11pt,paspconf,psfig]{article}

\begin{document}

\title{Globular Cluster Halos around the brightest Fornax Ellipticals}
\author{Markus Kissler-Patig}
\affil{UCO/Lick Observatory, University of California, Santa Cruz, CA
96054, USA}

\section{What it is all about}

Our goal is to understand the globular clusters surrounding the brightest
galaxies in the Fornax galaxy cluster, and relate their properties to
the formation history of the galaxies and the galaxy cluster. In general, 
we need to understand the presence of
globular cluster sub--populations around the brightest galaxies (and their
apparent absence around the less luminous ellipticals, Kissler-Patig 1997). 
In particular, the
extreme over--abundance of globular clusters around the central cD galaxy
is still unexplained. 

Fornax is a relatively poor galaxy cluster at a distance of $\simeq 19$ Mpc
(assumed throughout the following).
The properties of the brightest early--type galaxies (taken from the RC3,
Kissler-Patig et al.~1997a,b, and Forbes et al.~1997) are summarized below, 
including the number of globular
clusters (N$_{\rm GC}$), and the specific frequency (S$_N$).
NGC 1399, a giant cD galaxy, sits in the center of the cluster
gravitational potential,
while NGC 1404 and NGC 1380 are at projected distances of 50 kpc and 200 kpc to 
the SE and NW respectively.
\begin{center}
\begin{tabular}{ccccll}
\tableline
Name & M$_{V_T}$[mag] & B$-$V [mag]& Vel$_{\rm opt}$ [km s$^{-1}$] &  N$_{\rm GC}$ & S$_N$ \\
\tableline
NGC 1399 & $-21.8$ & 0.96 & 1447 & $\simeq 6000$ & $\simeq 11$ \\
NGC 1380 & $-21.5$ & 0.94 & 1841 & $\simeq 600$ & $\simeq 2$ \\
NGC 1404 & $-21.4$ & 0.97 & 1929 & $\simeq 750$ & $\simeq 2$ \\
\tableline
\end{tabular}
\end{center}

\section{The globular cluster systems of NGC 1399, NGC 1380, and NGC 1404}

$\bullet$ Qualitatively, NGC 1399, NGC 1380 and NGC 1404
have very similar globular cluster populations (Kissler-Patig et
al.~1997a,b, Forbes et al.~1997). All have bi--modal
globular cluster color distributions and have about the 
same number of red and blue globular clusters. The properties of these
sub--populations were investigated spectroscopically in NGC 1399 
(Kissler-Patig et al.~1998a), with
extremely deep 3--color photometry in NGC 1380 (Kissler-Patig et al.~1997b)
and with 2--color HST photometry in NGC 1404 (Forbes et al.~1997). The blue 
globular clusters show
very similar properties to the Milky Way halo globular clusters (similar
age and metallicity, spherically distributed and extended), and could
constitute a {\it halo}. On the other hand, the red globular clusters are
more metal rich (comparable to, maybe slightly richer than, the Milky Way
disk/bulge globular clusters). There is some evidence that the
latter could be a few gigayears younger than the former, but both 
sub--populations are old, and did {\it not} form in a late ($z<1$) merger.
This is not really surprising since
hierachical clustering models predict (e.g.~Kauffmann 1996) and stellar 
populations indicate (e.g.~Bender 1997), that these galaxies formed at higher 
redshifts. However, 
both NGC 1399 and probably NGC 1380 have a small number of {\it very red} 
globular clusters, that
could be the product of a more recent interaction. Unfortunatly, both
stellar population synthesis models and observations are not yet accurate 
enough to pin down their ages with any certainty.

$\bullet$ Quantitatively, the globular clusters in NGC 1399, NGC 1380
and NGC 1404 differ dramatically. NGC 1399 is only 0.3 magnitudes brighter
than NGC 1380, and only 0.4 mag brighter than NGC 1404. However, NGC 1399
hosts $\simeq 6000$ globular clusters (Kissler-Patig et al.~1997a, Forbes
et al.~1997), while NGC 1380 hosts only
$\simeq 600$ (Kissler-Patig et al.~1997b) and NGC 1404 $\simeq 750$ (Forbes
et al.~1997, Richtler et al.~1992). In terms of specific
frequency, NGC 1399 has a S$_N\simeq
11$, while both NGC 1380 and NGC 1404 have S$_N\simeq2$. The typical value (with
little scatter) for the low-luminosity early--type galaxies (NGC
1374, NGC 1379, NGC 1387, NGC 1427) is S$_N\simeq 3$ (Kissler-Patig et
al.~1997a).
{\it The globular clusters around NGC 1399 are not different from the ones
around NGC 1380 and NGC 1404, but are more numerous by about a factor
of 10.}

\section{Kinematics in NGC 1399 and Fornax}

A sample of 76 radial velocities for globular clusters around NGC 1399 was
compiled from (Grillmair et al.~1994, Minniti et al.~1998, Kissler-Patig et
al.~1998a). We found no significant correlation of velocity or 
velocity dispersion of the globular clusters with position, position angle, 
color, or radius, either in the whole sample or in any sub--sample. In particular,
we could not find any rotation, nor are the
``high velocity'' clusters clearly associated with either NGC 1380 or NGC
1404 (see Kissler-Patig et al.~1998b for the detailed analysis).

However, the velocity dispersion of the globular 
clusters (and any sub--sample!) is high ($400\pm40$ km  s$^{-1}$).
The figure below shows the velocity
dispersion of several components: triangles show the velocity dispersion of
the stars around NGC 1399 (Franx et al.~1989, Bicknell et al.~1989), stars 
mark the velocity dispersion of the planetary nebulae (Arnaboldi et
al.~1994), squares show the velocity dispersion of the cluster
galaxies (Den Hartog \& Katgert 1996, Ferguson 1989), and circles show the 
velocity of the globular clusters (solid)
and various globular cluster sub--samples (open, see Kissler-Patig et
al.~1998b for a detailed description).
The globular clusters have a velocity dispersion that contrasts with
that of the stellar component of NGC 1399, and seems rather similar
to that of the cluster galaxies. {\it The globular clusters appear to
feel the cluster potential, rather than the potential of NGC 1399} (as
already noticed by Grillmair et al.~1994).
\begin{figure}
\hskip 3cm \psfig{figure=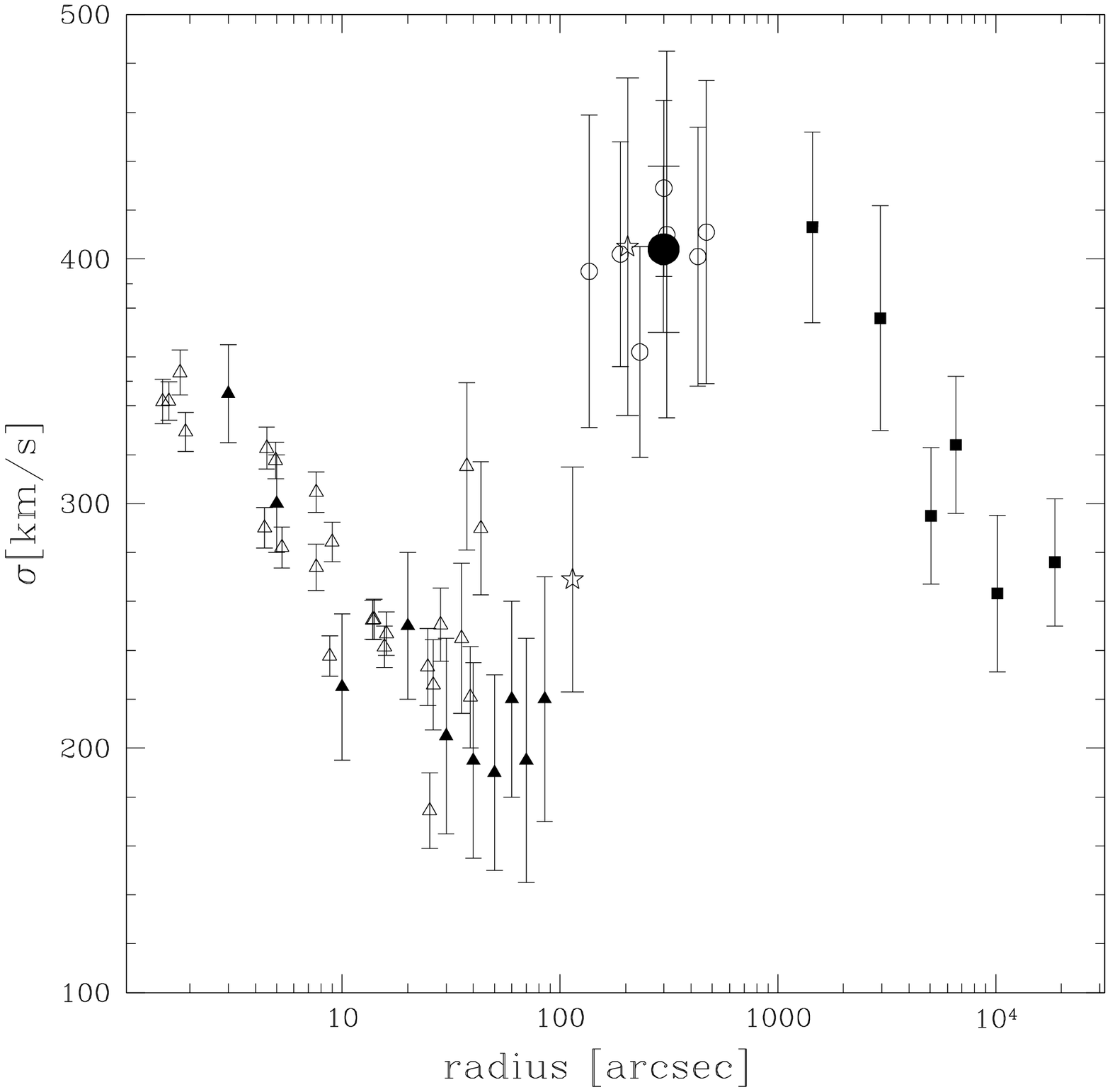,height=8cm,width=8cm
,bbllx=33mm,bblly=67mm,bburx=188mm,bbury=225mm}
\end{figure}

\section{Connecting the pieces of the puzzle}

The globular cluster system of NGC 1399 does not differ qualitatively from
the systems of NGC 1380 and NGC 1404, the second and third brightest galaxies in
Fornax, but hosts about 10 times more clusters. {\it The same mechanisms
seem to be at the origin of the globular clusters around all three galaxies}.
The only clear difference
between NGC 1399 and the two other galaxies, is its position in the very
center of the galaxy cluster. The over--abundance of globular clusters seem
therefore to be linked to the preferential position.
Furthermore, the kinematics of the globular clusters differ from those of
the stellar component, but ressemble those of the cluster galaxies.
{\it A large fraction of the globular clusters
seen around NGC 1399 seem to be associated with the galaxy cluster, rather
then with the galaxy itself.}

There are two possible explanation for the presence of the globular clusters 
in the center of the galaxy cluster: {\it a)} They might have formed in the
center of the galaxy cluster (e.g.~Balkeslee 1996), {\it b)} they might have been
accreted/stripped from the other cluster galaxies (Muzzio 1987, White 1987). At this point,
we cannot clearly distinguish between the two scenarios. But we note that there
is approximately the same number of blue and red globular clusters around 
NGC 1399. If the red (disk/bulge like) clusters are associated with the 
formation of the stellar component, and thus their number expected to scale 
with the galaxy size, then the excess of red clusters around NGC 1399 is
too large to be explained by formation in situ, and 
{\it stripping/harrassement must have played a role}. Stripping is further 
supported by the similarity of the globular clusters around the three galaxies.	
If stripping was the dominant
process, we expect ``tails'' of globular clusters towards the other
galaxies. These tails should be detectable in large photometric and 
spectroscopic studies.

\acknowledgments
It is a pleasure to thank J.P.Brodie, M.Della Valle, D.A.Forbes, P.Goudfooij,
C.J.Grillmair, M.Hilker, J.P.Huchra, L.Infante, S.Kohle, G.Meylan,
D.Minniti, H.Quintana, T.Richtler, L.L.Schroder, J.Storm, who  collaborated 
on the various parts of this work.


\begin{references}
\reference Arnaboldi, M., Freeman, K.C., Hui, X., Capaccioli, M. \& Ford, H.
1994, ESO Messenger, 76, 40
\reference Bender R., 1997, in ``The Nature of Elliptical Galaxies'', 
ASPCS Vol.116, eds. M.Arnaboldi, G.S.Da Costa, P.Saha, p.11
\reference Bicknell, G.V., Carter, D., Killeen, N., Bruce, T. 1989, ApJ, 336,
639
\reference Blakeslee J.S., 1996 PhD thesis, Massachusetts Institute of
Technology
\reference Den Hartog, R. \& Katgert, P. 1996, MNRAS, 279, 349
\reference Ferguson, H.C., 1989, AJ, 98, 367
\reference Forbes, D.A., Grillmair, C.J., Williger, G.M., et al. 1997, MNRAS, in
press
\reference Franx, M., Illingworth, G. \& Heckman, T. 1989, ApJ, 344, 613
\reference Grillmair, C.J., Freeman, K.C., Bicknell, G.V., et al., 1994, ApJ,
422, L9
\reference Kauffmann G., 1996, MNRAS 281, 487 
\reference Kissler-Patig, M. 1997, A\&A 319, 83
\reference Kissler-Patig, M., Kohle, S., Hilker, M., et al. 1997a, A\&A 319, 470
\reference Kissler-Patig, M., Richtler, T., Storm, J., Della Valle, M. 1997b, A\&A, in press 
\reference Kissler-Patig, M., Brodie, J.P., Schroder, L.L., et al. 1998a, AJ
January issue
\reference Kissler-Patig M., et al., 1998b, AJ in preparation
\reference Minniti, D., Kissler-Patig, M., Goudfrooij, P., Meylan, G.
1998, AJ January issue
\reference Muzzio, J.C. 1987, PASP, 99, 245
\reference Richtler T., Grebel E.K., Domg\"orgen H., Hilker M., Kissler M.,
1992, A\&A, 264, 25 
\reference White R.E. III, 1987, MNRAS, 227, 185
\end{references}
\end{document}